\documentclass[11pt]{amsart}

\usepackage{amsthm,amsmath,amssymb}
\usepackage[foot]{amsaddr}
\usepackage{graphicx}
\usepackage{natbib}
\usepackage{tikz}
\usetikzlibrary{fadings}
\tikzfading[name=fade down,top color=blue!0, bottom color=red!100]
\tikzfading[name=fade up,top color=red!100, bottom color=blue!0]

\usepackage[colorlinks=true,linkcolor=red,citecolor=blue,urlcolor=cyan]{hyperref}

\newtheorem{proposition}{Proposition}[section]

\usepackage{fancyhdr}
\title{Regression analysis of \\ unmeasured confounding}

\author{Brian Knaeble}
\address{Department of Mathematics, Utah Valley University, Orem, UT}
\email{bknaeble@uvu.edu}

\author{Braxton Osting}
\address{Department of Mathematics, University of Utah, Salt Lake City, UT}
\email{osting@math.utah.edu}

\author{Mark Abramson}
\address{Department of Mathematics, Utah Valley University, Orem, UT}
\email{mark.abramson@uvu.edu}

\date{\today}
\keywords{Propensity, Correlation, Model uncertainty}

\begin{document}
\maketitle

\begin{abstract}
When studying the causal effect of $x$ on $y$, researchers may conduct regression and report a confidence interval for the slope coefficient $\beta_{x}$. This common confidence interval provides an assessment of uncertainty from sampling error, but it does not assess uncertainty from confounding. An intervention on $x$ may produce a response in $y$ that is unexpected, and our misinterpretation of the slope happens when there are confounding factors $w$.  When $w$ are measured we may conduct multiple regression, but when $w$ are unmeasured it is common practice to include a precautionary statement when reporting the confidence interval, warning against unwarranted causal interpretation. 
If the goal is robust causal interpretation then we can do something more informative.  Uncertainty in the specification of three confounding parameters can be propagated through an equation to produce a confounding interval.  Here we develop supporting mathematical theory and describe an example application.  Our proposed methodology applies well to studies of a continuous response or rare outcome.  It is a general method for quantifying error from model uncertainty.  Whereas confidence intervals are used to assess uncertainty from unmeasured individuals, confounding intervals can be used to assess uncertainty from unmeasured attributes. 
\end{abstract}

\section{Introduction}
\label{int}
Causal inference from observational data is challenging and controversial \citep{Ding15}.  
Confounding bias can be eliminated, theoretically, using causal graphs to select an admissible set of covariates for adjustment \citep{Pearl09}, but this relies on a ``web of assumptions'' \citep{Rosecritique}.  
Confounding bias can be assessed with sensitivity analysis and propensity scores \citep{Rosenbaum10}, assuming estimates of propensity for each individual.  
In this paper, we show that when an admissible set of covariates is unmeasured and individual estimates for propensity are unavailable, it is still possible to conduct causal inference. Our introduced methodology allows us to bound an adjusted estimate without requiring individual observations on the confounders. 

For simplicity and concreteness we focus on the model
\begin{equation}
\label{model}
y=\beta_{0|w}+\beta_{x|w}x+\beta_1w_1+\cdots+\beta_pw_p+\varepsilon.
\end{equation}
We have $n$ observations on $(x,y)$, where $x$ has recorded treatment or exposure values and $y$ has measured the resulting response or outcome.  The unmeasured confounding set $w=\{w_1,\cdots,w_p\}$ may contain indicator variables for homogeneous groups or higher order interaction terms.
The joint error distribution is assumed to be consistent with the principle of least-squares.  We can not fit the model since $w$ is unmeasured, and we should not fit the reduced model \[y=\beta_0+\beta_x x+\tilde{\epsilon}\] because the unadjusted slope-coefficient $\beta_x$ may differ greatly from the adjusted slope-coefficient $\beta_{x|w}$ \citep{KD}.  We can, however, obtain $\beta_{x|w}$ from the parameters 
\[R^2_{wx} \qquad  R^2_{wy}, \quad \textrm{and} \quad \rho_{\hat{x}\hat{y}}.\]

The coefficient of determination
$R^2_{wx} = \frac{ | \hat x - \bar x|^2 }{ |x - \bar x|^2}$ 
is the proportion of variation in $x$ explained by $w$ using the model $x=\alpha_0+\alpha_1w_1+\cdots+\alpha_pw_p$ 
fit with least-squares. 
Here, $\bar x = \frac{1}{n} \sum_{i=1}^n x_i$ is the mean and $\hat x=\alpha_0+\alpha_1w_1+\cdots+\alpha_pw_p$ are the fitted values.  
Similarly, the coefficient of determination 
$R^2_{wy} = \frac{ | \hat y - \bar y|^2} { | y - \bar y |^2 } $ 
is the proportion of variation in $y$ explained by $w$ using the model $y=\gamma_0+\gamma_1w_1+\cdots+\gamma_pw_p$ fit with least-squares.  Here, $\bar y = \frac{1}{n} \sum_{i=1}^n y_i$ is the mean and $\hat y=\gamma_0+\gamma_1w_1+\cdots+\gamma_pw_p$ are the fitted values.
The coefficient  
$\rho_{\hat{x}\hat{y}}= \frac{ \left\langle\hat{x} - \bar{\hat x},\hat{y}-\bar{ y} \right\rangle }{\left|\hat{x} -\bar{x}\right| \left|\hat{y} - \bar{\hat y}\right|}$ is the Pearson correlation coefficient between the two vectors of fitted values.  
\begin{proposition}
\label{formula}
Let $\rho_{xy} \in [-1,1]$ be the measured correlation coefficient for $x$ and $y$, and let
$\sigma_y/\sigma_x>0$ be the measured ratio of standard deviations for $y$ and $x$. 
With $R^2_{wx}$, $R^2_{wy}$, and $\rho_{\hat{x}\hat{y}}$ as described above, the adjusted slope-coefficient $\beta_{x|w}$ satisfies 
\begin{equation}
\label{eq1}
\beta_{x|w}=\frac{\sigma_y}{\sigma_x}\frac{\rho_{xy}-R_{wx}R_{wy}\rho_{\hat{x}\hat{y}}}{1-R_{wx}^2}.
\end{equation}
\end{proposition}

Proposition \ref{formula} can be derived from a proof found in the appendix of \citet{KD}.  We provide an alternative proof using projection matrices in Appendix~\ref{ap:1}.  Within \eqref{eq1} the terms $R_{wx}R_{wy}\rho_{\hat{x}\hat{y}}$ and $R_{wx}^2$ can be rewritten as $\rho_{\hat{x}x}\rho_{\hat{x}y}$ and $\rho^2_{\hat{x}x}$ respectively (c.f. \citet{Frank}), but there are advantages to our factorized formulation.  $R_{wx}$ and $R_{wy}$ are square roots of coefficients of determination and therefore monotonic in $p$, the number of predictors of $w$.  Also, the factors $R_{wy}=\rho_{\hat{y}y}$ and $\rho_{\hat{x}\hat{y}}$ are more intuitive than their product $\rho_{\hat{x}y}$.  We develop intuition for $\rho_{\hat{x}\hat{y}}$ within the context of our case study in Section~\ref{app}.

The $3$-tuple $(R^2_{wx},R^2_{wy},\rho_{\hat{x}\hat{y}})$ provides insight into unmeasured confounding in a way similar to how epidemiologists use association parameters during categorical sensitivity analysis or bias analysis.  They often make use of information on the prevalence of an unmeasured confounder, its association with treatment or exposure, and its effect on the outcome \citep[p. 548]{Mac}.  \citet{SAWA} have used risk ratios in a condition supporting the E-value \citep{EV}.  \citet{Lee11} has a condition utilizing risk differences and odds ratios.  The performance of conditions has been assessed with simulations \citep{KC}, and  conditions originally derived in the continuous setting have performed well.


For continuous sensitivity analysis \citet{Frank} has developed an index to bound the impact of confounders, and \citet{HHH} have established reference points for speculation about omitted confounders.  \citet{KD} have shown how to determine the sign of $\beta_{x|w}$ when $R_{wx}R_{wy}<|\rho_{xy}|$.  
Here we improve their result by showing how to capture $\beta_{x|w}$ within an interval.  
This \emph{confounding interval} is computed using an algorithm described in Section~\ref{meth}. 
Supporting proofs are provided in the appendix.  In Section~\ref{app} we describe an example case study emphasizing uncertainty from unmeasured attributes over uncertainty from unmeasured individuals (c.f. \citet{RRC}). In Section \ref{disc} we discuss extensions and limitations of our methodology while emphasizing connections with topics of interest in causality, the Bayesian paradigm, categorical sensitivity analysis, propensity scores, and high-dimensional data analysis.

\section{Methods} \label{meth}
We specify six interval endpoints $\{l_x^2,u_x^2,l_y^2,u_y^2,l_{\hat{x}\hat{y}},u_{\hat{x}\hat{y}}\}$ determining the constraints 
\begin{subequations} \label{e:intervalBounds}
\begin{align}
\label{e:intervalBoundsa}
 0&\leq l_x^2\leq R^2_{wx}\leq u_x^2<1 \\
\label{e:intervalBoundsb}
0&\leq l_y^2\leq R^2_{wy}\leq u_y^2<1 \\
\label{e:intervalBoundsc}
-1&\leq l_{\hat{x}\hat{y}}\leq \rho_{\hat{x}\hat{y}}\leq u_{\hat{x}\hat{y}}\leq 1.
\end{align}  
\end{subequations}
Not all tuples $(R^2_{wx},R^2_{wy},\rho_{\hat{x}\hat{y}})$ that satisfy the constraints are realistic.  
To characterize realistic tuples we need a fourth constraint.  
The following proposition is proven in Appendix~\ref{ap:2}.
\begin{proposition}
\label{prop2}
The tuple of statistics $(\rho_{xy}, R^2_{wx}, R^2_{wy}, \rho_{\hat{x}\hat{y}})$ can arise from actual data only if 
\begin{equation}\label{nonline}
\rho_{\hat{x}\hat{y}}\in [\alpha_-,\alpha_+], 
\qquad \textrm{where} \quad 
\alpha_\pm=\frac{\rho_{xy} \pm \sqrt{1-R^2_{wx}} \sqrt{1-R^2_{wy}} } {R_{wx} R_{wy}}.
\end{equation} 
Conversely, if  $n>p+2$ and the coefficients satisfy 
$$
\rho_{xy} \in (-1,1), \qquad 
R_{wx}^2 \in [0,1), \qquad
R_{wy}^2 \in [0,1), \quad and \quad
\rho_{\hat{x}\hat{y}} \in [\alpha_-,\alpha_+], 
$$
then we can construct data for these coefficients which realize the tuple 
$(\rho_{xy}, R^2_{wx}, R^2_{wy}, \rho_{\hat{x}\hat{y}})$.
\end{proposition}

\begin{figure}[b]
\centering
\includegraphics[width=0.6\textwidth]{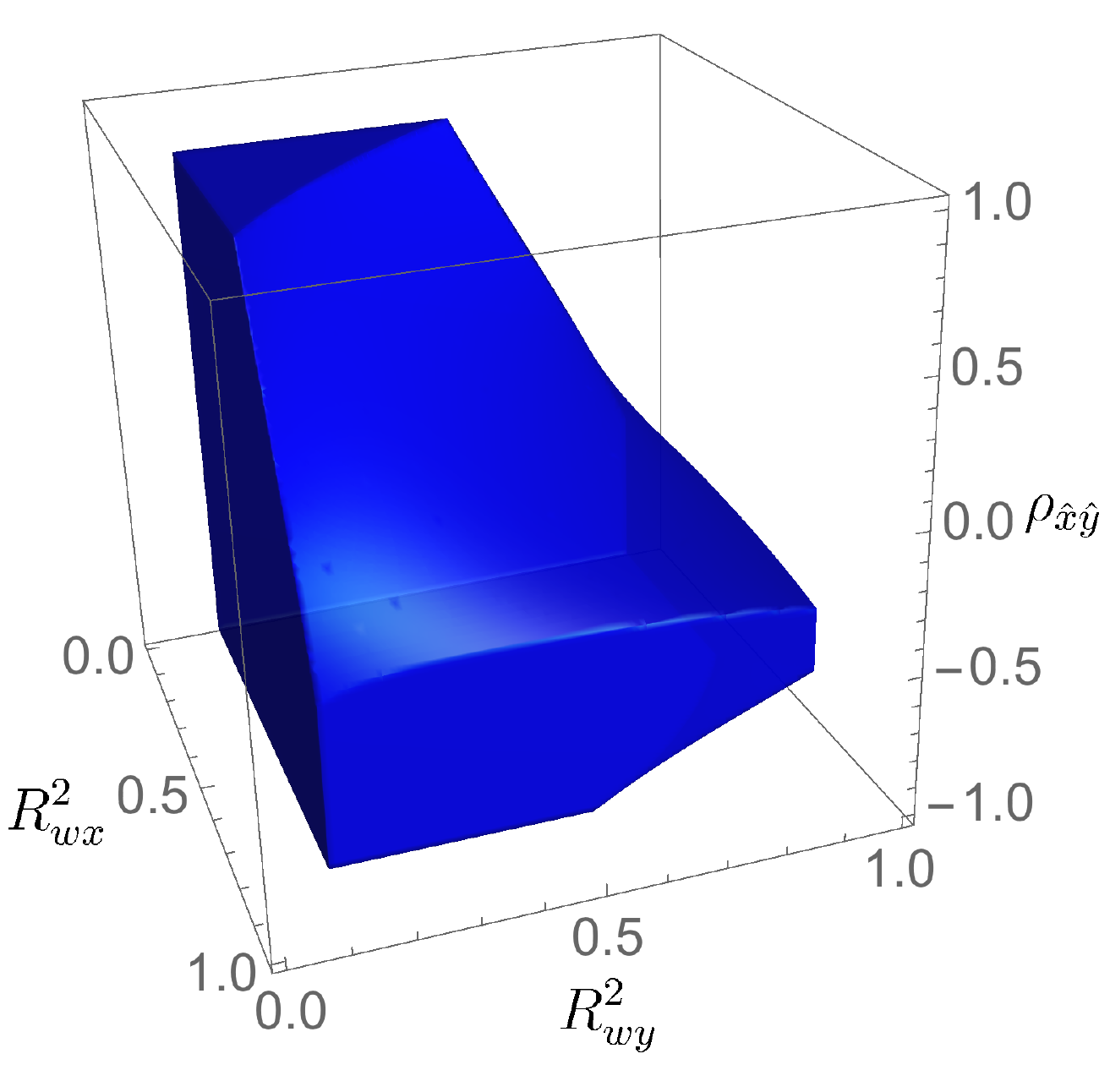}
\caption{An illustration showing one of many possible configurations for the constraint set, $\Omega$, with specifications 
$\rho_{xy} = -0.4$, 
$l_x^2 = l_y^2 = 0.1$, 
$u_x^2 = u_y^2 = 0.9$, 
$l_{\hat{x}\hat{y}} = -0.9$, and
$u_{\hat{x}\hat{y}} = 0.9$. }
\label{f:Omega}
\end{figure}

The intersection $\Omega$ of the four constraints given in \eqref{e:intervalBounds} and \eqref{nonline} is assumed to be nonempty.  An example of $\Omega$ for one choice of parameters is given in Figure~\ref{f:Omega}; the parameters are specified in the caption.  We refer to $\Omega$ as the feasible set.  A point $(R_{wx}^2,R_{wy}^2,\rho_{\hat{x}\hat{y}})$ is said to be \emph{feasible} if it is an element of $\Omega$. 

Given $(\rho_{xy},\sigma_y/\sigma_x)$, the function 
\begin{equation} 
\label{e:confInt}
\beta_{x|w}\colon 
(R_{wx}^2,R_{wy}^2,\rho_{\hat{x}\hat{y}})
\mapsto \frac{\sigma_y}{\sigma_x}\frac{\rho_{xy}-\sqrt{R^2_{wx}} \sqrt{R^2_{wy}}\rho_{\hat{x}\hat{y}}}{1-R_{wx}^2}
\end{equation}
is continuous on $\Omega$.  Since the function is continuous and $\Omega$ is connected, by the intermediate value theorem, $\beta_{x|w}(\Omega)$ is an interval.  By the Weierstrass extreme value theorem, the interval is closed and we write 
\begin{equation} \label{e:CI}
\beta_{x|w}(\Omega)=[l,u], 
\qquad \textrm{where} \quad  
l=\min_{\Omega} \ \beta_{x|w} \quad 
\textrm{and} \qquad
u=\max_{\Omega} \ \beta_{x|w}.
\end{equation}
We refer to $[l,u]$ as a \emph{confounding interval}. 

Computation of a confounding interval requires solutions to the non-convex, constrained optimization problem in \eqref{e:CI}, {\it i.e.}, minimizing and maximizing $\beta_{x|w}$, over the feasible set $\Omega$.  We have developed an algorithm for computing any confounding interval from input parameters \[\{\rho_{xy},\sigma_y/\sigma_x;l_x^2,u_x^2,l_y^2,u_y^2,l_{\hat{x}\hat{y}},u_{\hat{x}\hat{y}}\}.\]
The algorithm computes exact solutions in negligible run time.  It is based on the following proposition.
\begin{proposition}
\label{prop3}
Let $q_\pm^2(a,b,c)$ denote the square of the two solutions to the quadratic equation $ax^2+bx+c=0$, {\it i.e.}, $q_\pm^2(a,b,c) = \left( \frac{-b}{2a} \pm \frac{\sqrt{b^2 - 4 a c}}{2a}\right)^2$.
Let $S \subset \mathbb R^3$ be the discrete set of points $(R_{wx}^2,R_{wy}^2,\hat{\rho}_{xy})$ that are feasible and of one of the following forms: 
\begin{subequations}
\begin{align}
\label{firstset}
&\left(q_\pm^2(-b_yb_{\hat{x}\hat{y}},2\rho_{xy},-b_yb_{\hat{x}\hat{y}}), \ b_y^2, \ b_{\hat{x}\hat{y}}\right) \\
\label{secondset}
& \left((\rho_{xy}+ 1)/(b_{\hat{x}\hat{y}}+ 1), \ (\rho_{xy}+ 1)/(b_{\hat{x}\hat{y}}+ 1), \ b_{\hat{x}\hat{y}}\right) \\
\label{secondsetb}
& \left((\rho_{xy}- 1)/(b_{\hat{x}\hat{y}}- 1), \ (\rho_{xy}- 1)/(b_{\hat{x}\hat{y}}- 1), \ b_{\hat{x}\hat{y}}\right) \\
\label{thirdset}
& \left(b_x^2, \ b_y^2, \ b_{\hat{x}\hat{y}}\right) \\
\label{fourthset}
& \left(b_x, \ b_y, \ (\rho_{xy}\pm\sqrt{1-b_x^2}\sqrt{1-b_y^2})/(b_xb_y)\right) \\
\label{fifthset}
& \left(b^2_x, \ q_\pm^2(b_x^2b_{\hat{x}\hat{y}}^2+1-b^2_x, \ -2b_xb_{\hat{x}\hat{y}}\rho_{xy}, \ b_x^2-1+\rho_{xy}^2), \ b_{\hat{x}\hat{y}}\right)\\ 
\label{sixthset}
& \left(q_\pm^2(b_x^2b_{\hat{x}\hat{y}}^2+1-b^2_x, \ -2b_xb_{\hat{x}\hat{y}}\rho_{xy}, \ b_x^2-1+\rho_{xy}^2), \ b_y^2, \ b_{\hat{x}\hat{y}}\right),
\end{align}
\end{subequations}
where 
$b_x^2\in \{l_x^2,u_x^2\}$,
$b_y^2\in\{l_y^2,u_y^2\}$, and
$b_{\hat{x}\hat{y}}\in\{l_{\hat{x}\hat{y}},u_{\hat{x}\hat{y}}\}$.
Then 
\[\min_{S}(\beta_{x|w})=\min_{\Omega}(\beta_{x|w})=l
\qquad  \textrm{and} \qquad
\max_{S}(\beta_{x|w})=\max_{\Omega}(\beta_{x|w})=u.\]
\end{proposition}
A proof of Proposition~\ref{prop3} is given in Appendix~\ref{ap:3}.  The set $S$ is finite with cardinality $|S|\leq 88$ for all sets of parameter values.  Given input parameters the computational algorithm first determines $S$ and then computes $\min_{S}(\beta_{x|w})$ and $\max_{S}(\beta_{x|w})$, which are the endpoints of the desired confounding interval according to Proposition \ref{prop3}.  
Python and R implementations of this algorithm are provided at the first author's github page \citep{github}.

%


\section{Application}
\label{app}
\citet{Eskenazi13} studied $n=248$ children and found an association between in utero PBDE exposure (log transformed) and follow-up IQ at 7 years of age.  Regression of $y=\textrm{IQ}$ on $x=\textrm{PBDE exposure}$ produces a slope estimate $\hat{\beta}_{xy}=-4.48$ with a standard error $se(\hat{\beta}_{xy})=2.71$.  While there is some uncertainty about whether this finding is statistically significant, there is also uncertainty about whether causal interpretation is warranted.  For reference we record the standard deviations $\sigma_x=0.34$ and $\sigma_y=14.60$ ($\sigma_y/\sigma_x=42.94$) and the correlation $\rho_{xy}=-0.11$. 


We can use potential outcomes $y_x$ to define non-confounding or ignorable treatment or exposure assignment \citep{Rose83}.  Causal inference is warranted when conditional on some covariate set the potential outcomes are independent of treatment: $(y_x\perp \!\!\! \perp x)|w$.  The assumption of ignorable treatment assignment can be made more believable by conditioning on as many pretreatment covariates as possible \citep[p. 76]{Rubin09,Rose02}.  Eskenazi et al. adjusted for $\{$sex, mothers score on a vocabulary test, spoken language, maternal years spent living in the US, parity, and exposure to environmental tobacco smoke$\}$.  Their adjusted estimate was $\hat{\beta}_{xy|adj.}=-5.60$.

It is possible for an adjusted estimate to have amplified bias \citep{Ding15}.  There are (non statistical, causal) methods for selecting an admissible set of covariates for adjustment \citep{Pearl09}.  If the admissible set is unmeasured some insight can be gained by using our algorithm (see Section~\ref{meth}).  We can construct a confounding interval to assess uncertainty from unmeasured confounding by any set of confounders.  To demonstrate our proposed methodology we consider unmeasured confounding by diet, simplified as $w$: $\{$fat consumption, protein consumption$\}$.


\citet{VS11} recommend adjustment for any covariate that causes exposure or the outcome.  There are reasons to suspect diet as a cause of PBDE exposure \citep{PBDEfood, CDC}.  Suppose for illustrative purposes that we bound diet's coefficient of determination for PBDE exposure as follows: $10\%\leq R_{wx}^2\leq 50\%$.  Diet may also be a causal factor for IQ \citep{DietIQ,ProtIQ}.  Suppose again for illustrative purposes that we bound diet's coefficient of determination for IQ as follows: $0\%\leq R_{wy}^2\leq 20\%$.  Based on this information alone we may apply our optimization algorithm (see Section~\ref{meth}) and conclude $\beta_{x|w}\in [-36.60,17.71]$.  
Since the upper and lower bounds for $\rho_{\hat{x}\hat{y}}$ were left unspecified we have set $l_{\hat{x}\hat{y}}=-1$ and $u_{\hat{x}\hat{y}}=1$ by default.  

Tighter bounds on $\beta_{x|w}$ are possible through careful specification of $l_{\hat{x}\hat{y}}$ and $u_{\hat{x}\hat{y}}$ to bound $\rho_{\hat{x}\hat{y}}$.  We describe two hypothetical situations to improve reader intuition for $\rho_{\hat{x}\hat{y}}$.  In each scenario we have within $w$ an $\hat{x}$ largely determined by the first variable, fat consumption \citep{PBDEfood}.  On a population that consumes fish we may have within $w$ a $\hat{y}$ also largely determined by the first variable, (beneficial) fat consumption \citep{DietIQ}, resulting in $\rho_{\hat{x}\hat{y}}$ near $1$.  
On a population that does not consume fish we may have within $w$ a $\hat{y}$ now largely determined by the second variable, protein consumption \citep{ProtIQ}, resulting in $\rho_{\hat{x}\hat{y}}$ near $0$. 

\begin{figure}[h]
\centering
\includegraphics[width=0.8\textwidth]{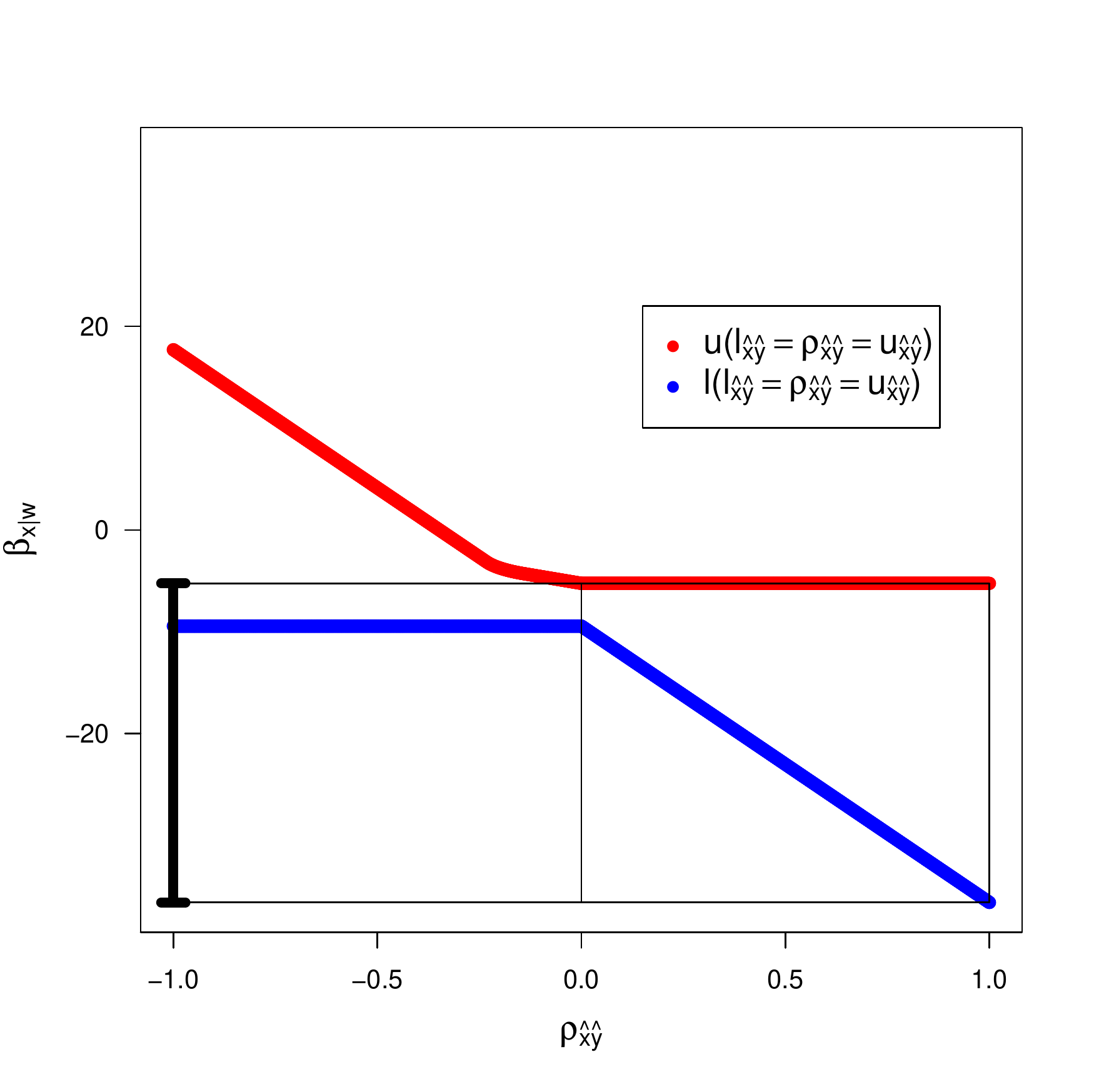}
\caption{A plot showing the dependence of the confounding interval $[l,u]$ on user specified $(l_{\hat{x}\hat{y}},u_{\hat{x}\hat{y}})$ given $\rho_{xy}=-0.11$, $\sigma_y/\sigma_x=42.94$, $l_x^2=10\%$, $u_x^2=50\%$, $l_y^2=0\%$, and $u_y^2=50\%$, e.g. $\beta_{x|w}\in [-36.60,-5.25]$ if $\rho_{\hat{x}\hat{y}}\in [0,1]$.} 

\label{f:VaryingRho}
\end{figure}

Negative confounding \citep{Choi} occurs when $\rho_{\hat{x}\hat{y}}<0$.  If for illustrative purposes we rule out negative confounding and specify $0\leq \rho_{\hat{x}\hat{y}}\leq1$ in addition to our previously specified $0.1\leq R_{wx}^2\leq 0.5$ and $0\leq R_{wy}^2\leq 0.2$ then another application of the optimization algorithm produces the tighter bound $\beta_{x|w}\in [-36.60,-5.25]$.  This interval can be seen in black on the lower left portion of the plot in Figure~\ref{f:VaryingRho}. The plot shows a rough dependence of $\beta_{x|w}$ on $\rho_{\hat{x}\hat{y}}$.  Open source software for making similar plots is provided at the first author's github page \citep{github}.  

\section{Discussion} \label{disc}
We have gained insight into unmeasured confounding using coefficients of determination and correlation between fitted values.  Uncertainty of these coefficients can be propagated through Equation \ref{eq1} to produce a confounding interval.  The details of this methodology have been described in Section \ref{meth}, and an example case study application has been described in Section \ref{app}.  In this section, we elaborate by describing various extensions and some limitations while emphasizing connections to related topics of interest.

\begin{figure}[h]
\centering
\includegraphics[width=0.8\textwidth]{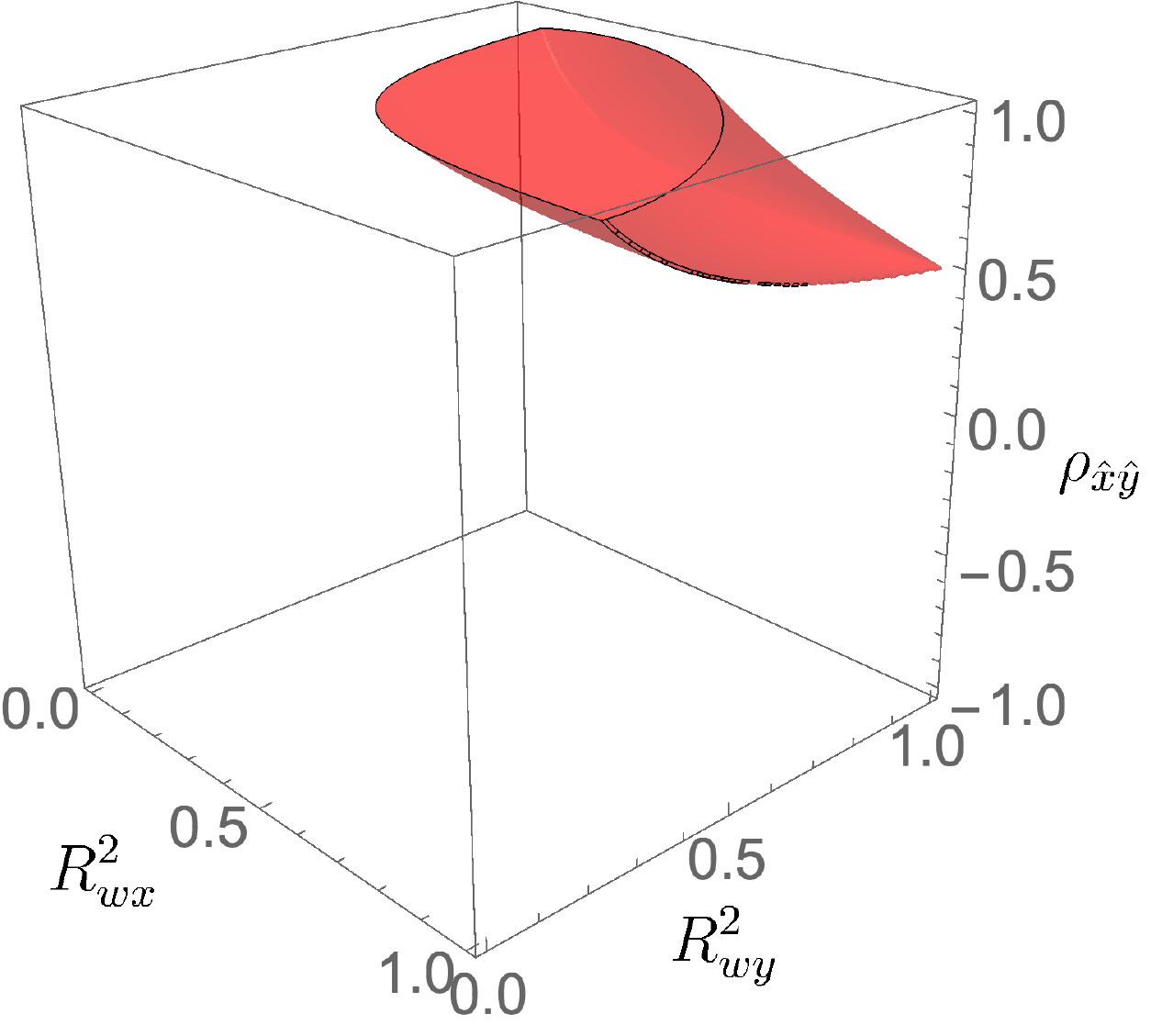}
\caption{We have used \eqref{nonline} and \eqref{e:confInt} to plot the subset of $(R_{wx}^2,R_{wy}^2,\rho_{\hat{x}\hat{y}})$-values that satisfy $\rho_{\hat{x}\hat{y}}\in [\alpha_-,\alpha_+]$ as in \eqref{nonline} and $\beta_{x|w}\not\in [.2,\infty]$ given $\rho_{xy}=0.5$ and $\sigma_y/\sigma_x=1$.}
\label{f:InverseProb}
\end{figure}

\subsection{Necessary conditions}
To avoid assumptions \citep[p. 369]{SAWA} a researcher may seek conditions on $(R_{wx}^2,R_{wy}^2,\rho_{\hat{x}\hat{y}})$ that are necessary for any interpretation to explain away the observed association $\rho_{xy}$ \citep{Corn,Ding14}.  We could use \eqref{nonline} and \eqref{e:confInt} to determine the subset of $(R_{wx}^2,R_{wy}^2,\rho_{\hat{x}\hat{y}})$-values that are realizable from an actual $w$ and necessary for $\beta_{x|w}$ to be practically insignificant.  
This approach is illustrated in Figure \ref{f:InverseProb}.  
If the resulting subset is unreasonably extreme (i.e. inconsistent with subject matter knowledge) then we may infer from the $(x,y)$-data and supporting analysis that $\beta_{x|w}$ is practically significant for all $w$.  To avoid retrospective confirmation bias we recommend prospective definition of practical significance and reasonable $(R^2_{wx},R^2_{wy},\rho_{\hat{x}\hat{y}})$-values.     

\subsection{Bayesian paradigm}
\label{Bayes}
We have seen how $\rho_{\hat{x}\hat{y}}$ is important during assessment of uncertainty from unmeasured confounding.  Researchers may also bound $\rho_{\hat{x}y}=R_{wy}\rho_{\hat{x}\hat{y}}$ or even $\rho_{x\hat{y}}=R_{wx}\rho_{\hat{x}\hat{y}}$, and bounds on these quantities manifest as additional constraints on $(R_{wx}^2,R_{wy}^2,\rho_{\hat{x}\hat{y}})$.  More generally, researchers may specify an arbitrary subset of possible $(R_{wx}^2,R_{wy}^2,\rho_{\hat{x}\hat{y}})$-values.  On the basis of previously published literature a prior distribution on $(R_{wx}^2,R_{wy}^2,\rho_{\hat{x}\hat{y}})$ may be specified and transformed through Proposition~\ref{formula} to produce a distribution on $\beta_{x|w}$. 
Our methodology can thus be used to support reuse of externally valid experiments to facilitate causal inference from observational studies.  
Supplementary to classic use of confidence intervals in meta analyses, confounding intervals can be used to check for consistency between observational studies and controlled trials in a systematic review.   

\subsection{Causal inference}
An admissible $w$ is required for causal interpretation of a confounding interval, and toward this ideal analysts should consider all causes of variation in $x$ for inclusion within $w$ \citep{VS11}.  For instance, in the application study of Section \ref{app}, we could have included household use of electronics \citep{Horton,Ramani} in addition to diet.  Researchers can also adjust for covariates that cause $y$ \citep{VS11}, and if a covariate causes $y$ but not $x$ then it can be used to transform $y$ \citep{KnaebleAdj} before applying our method.  Some precaution is warranted when adjusting for confounders that are themselves not randomly assigned \citep[Section 2, Butterfly Bias]{Ding15}. \citet[Chapter 2]{Fisher} has written of randomization as the ``reasoned basis'' for causal inference \citep[p. 33]{Rosenbaum10}.  When there has been random assignment to $w$ we may then reason causally to bound $\rho_{\hat{x}\hat{y}}$.   

\subsection{Extensions and limitations}
We can repeatedly apply our methodology across the levels of a moderator variable to study effect modification or interaction effects \citep{Corraini,Vand}.  If subgroup sample sizes are small we may incorporate sampling distributions for $\rho_{xy}$ and $\sigma_y/\sigma_x$.  We do not recommend use of our methodology to assess uncertainty from unmeasured confounding of already-adjusted slope-coefficients, unless coefficients of determination and correlations between residual vectors remain intuitive.  Measurements on a subset $s\subset w$ of an admissible set of covariates can be used to estimate lower bounds $R^2_{sx}\leq R^2_{wx}$ and $R^2_{sy}\leq R^2_{wy}$.  We describe a technique for estimating upper bounds in Section \ref{hdd}.  We can not conduct thorough regression diagnostics when part of $w$ is unmeasured.  Also, we require a model that is linear in its parameters which precludes straightforward application during logistic regression \citep[Chapter 20]{Roth}.  Some modification is required before application during interrupted time series analysis \citep{Kontopantelis}.     

\subsection{Categorical predictors}
We do not require linearity of observed $(x,y)$ data.  Our main assumption is a causal effect $\beta_{x|w}$ that for some $w$ is linear, perhaps only locally, and perhaps only on some suitable subpopulation.  We automatically have a linear $\beta_{x|w}$ when $x$ is dichotomous categorical, e.g. when it is an indicator for a treatment or an exposure.  Then we can bound the magnitude of (the linear quantity) $R^2_{wx}$ using a (nonlinear) model of propensity \citep{Rose83} and a generalized $R^2$.  This framework is flexible enough to provide some support during causal interpretation of a comparison between two means.  For improved modeling we recommend use of indicator variables within $w$, c.f. \citet[Section 5.1]{Hastie}. When $w$ contains categorical variables they can be replaced with indicator variables in the standard way.  

\subsection{Categorical outcomes}
 When $y$ is dichotomous-categorical and the categories can be determined from a continuous latent model (as possible with say a Probit model) then we can estimate the latent $\rho_{xy}$ and $\sigma_y/\sigma_x$ values from the fitted parameters of the categorical model and proceed with the continuous analysis, or we can use least-squares estimates directly to estimate proportions. 
 If $y$ is rare and $x$ acts on (and only on) a fixed proportion of individuals then conditional rate differences may be roughly constant across the levels of a categorical $w$ and the adjusted (for $w$) rate difference can be approximated with $\beta_{xy|w}$.  If $y$ is common then inequalities \citep[Lemma 5.3]{KC} and approximations \citep{VRROR} provide some support for using regression coefficients to analyze unmeasured confounding of a contingency table \citep{KC}.

\subsection{Propensity scores}
\citet[p. 43]{Rose83} write that the propensity score is the coarsest balancing score. Hypothetically or ideally under their assumptions fine covariate data is no longer needed; coarse propensity is sufficient.  Here we go a step further, toward more coarseness, essentially collapsing rows (of individuals) in addition to columns (of attributes).  Under our assumptions propensity is no longer needed; the three statistics $R^2_{wx}$, $R^2_{wy}$, and $\rho_{\hat{x}\hat{y}}$ are sufficient, where $\hat{x}$ estimates propensity for treatment.  It is the correlation between propensity ($\hat{x})$ and treatment ($x$) that matters most (see the role of $R_{wx}=\rho_{\hat{x}x}$ in \eqref{eq1}).  This correlation is related to the distribution of propensities.  This correlation is close to zero when the propensities are nearly constant, and it is close to one when the propensities are near zero and one.  This correlation is determined by the distribution of propensities, given a sufficiently large sample.
\subsection{High-dimensional data} 
\label{hdd}
It can be difficult to interpret an observational study of high-dimensional data when there is model uncertainty \citep{Chat95}, and it may not be computationally feasible to fit every possible candidate model \citep{Patel}. Once a parsimonious model \citep{Occ} has been selected confounding intervals can be used to quickly assess the sensitivity of interpretations to any model extension within some space of regular extensions.  Since $R^2$ is monotonic in the number of predictors we can objectively set the constraints $u_x$ and $u_y$ by considering the largest set of predictors omitted from the parsimonious model and analyzing their coefficients of determination for $x$ and $y$ conditional on the set of covariates in the parsimonious model \citep{KD}. 

\subsection{Natural experiments}
We have an objective approach to assess uncertainty of residual confounding during causal analysis of high-dimensional observational data.  We also have a way to assess potential for unmeasured confounding using the parameters $R^2_{wx}$, $R^2_{wy}$, and $\rho_{\hat{x}\hat{y}}$. The parameter $R^2_{wx}$ gives the proportion of variation in $x$ that is due to $w$, and here it can be understood as a measure of departure from a randomized experiment (where $R^2_{wx}=0$) to a study with fully deterministic treatment assignment (where $R^2_{wx}=1$).   Some studies are quasi-experiments, and when variation in $x$ is mostly haphazard researchers may refer to the study as a natural experiment \citep[p. 5]{Rosenbaum10}.  This language for classifying studies can be made more precise with specification of bounds for $R_{wx}^2$, $R_{wy}^2$, and $\rho_{\hat{x}\hat{y}}$.  These bounds can then be transformed into confounding intervals for improved causal inference.

\section*{Acknowledgement}
We thank Tom Greene, Judy Ou, Jincheng Shen, and Yue Zhang for various suggestions and James Miles for his derivation of Proposition 2.1.


\clearpage 
\appendix
\section{Proof of Proposition~\ref{formula}} \label{ap:1}

Let $e\in \mathbb R^n$ denote the ones vector, $W = [e \mid w_1 \mid \cdots \mid w_p ]$,  
$$
P = W (W^t W)^{-1} W^t,  
\qquad \textrm{and} \qquad 
E = \frac{1}{n} e e^t.
$$ 
Note that $P$ and $E$ are projection matrices, {\it i.e.}, $P^2 = P$ and $E^2 = E$. Furthermore, since $W$ includes $e$ as a column, we have that $EP = PE = E$, $(I-P)(I-E) = (I-E) (I-P) = I-P$, and $P-E$ is also a projection matrix. 

We use $P$ and $E$ to write $\hat x = P x$, $\hat y = P y$, 
$$
\sigma_x^2 = \frac{1}{n} |(I-E)x|^2, 
\qquad \qquad 
\sigma_y^2 = \frac{1}{n} |(I-E)y|^2, 
$$
$$
\sigma_{\hat x}^2 = \frac{1}{n} |(I-E) P x|^2, 
\qquad \qquad 
\sigma_{\hat y}^2 = \frac{1}{n} |(I-E) P y|^2, 
$$
$$
R_{wx}^2 = \frac{\sigma_{\hat x}^2}{\sigma_{x}^2} = \frac{ |(I-E)Px|^2}{|(I-E)x|^2}, 
\qquad \qquad 
R_{wy}^2 = \frac{\sigma_{\hat y}^2}{\sigma_{y}^2} = \frac{ |(I-E)Py|^2}{|(I-E)y|^2},
$$
$$
\rho_{xy} = \frac{ \langle (I-E) x, (I-E) y \rangle}{ n \sigma_x \sigma_y}, 
\qquad \qquad 
\rho_{\hat x \hat y} = \frac{ \langle (I-E) P x, (I-E) P y \rangle}{ n \sigma_{\hat x} \sigma_{\hat y} },  
$$
and
$$
\rho_{(x-\hat x)(y- \hat y)} = \frac{ \langle (I-P) x, (I-P) y \rangle}{ |(I-P) x| \ |(I-P) y|}.
$$
Using the above expressions, we compute
\begin{align*}
n \sigma_x \sigma_y \rho_{xy}
&= \langle (I-E) x, (I-E) y \rangle \\
&= \langle (I-E) P x, (I-E) P y \rangle + \langle (I-P) x, (I-P) y \rangle \\
&= n \sigma_{\hat x} \sigma_{\hat y} \rho_{\hat x \hat y} + n \sigma_x \sigma_y \sqrt{1-R^2_{wx}} \sqrt{1-R^2_{wy}} \rho_{(x-\hat x)(y- \hat y)}.
\end{align*}
Dividing both sides by $n \sigma_x \sigma_y$, we obtain
\begin{equation}
\label{second} 
\rho_{xy} = R_{wx}R_{wy}\rho_{\hat{x}\hat{y}}+\sqrt{1-R_{wx}^2}\sqrt{1-R_{wy}^2} \rho_{(x-\hat{x})(y-\hat{y})}.
\end{equation}
Solving for $\rho_{(x-\hat{x})(y-\hat{y})}$, we obtain an expression for the partial correlation, 
\begin{equation}
\label{2b}
\rho_{(x-\hat{x})(y-\hat{y})} = \frac{ \rho_{xy} -  R_{wx}R_{wy} \rho_{\hat{x}\hat{y}}}{\sqrt{1-R_{wx}^2}\sqrt{1-R_{wy}^2}}. 
\end{equation}

We now consider our model from \eqref{model} rewritten as 
$$
y = \beta_{x|w} x + W \beta + \varepsilon, 
$$
where $\beta=(\beta_{0|w},\beta_1,\cdots,\beta_p)$.  Write $X = [x \mid W ]$ and $Q = X (X^t X)^{-1} X^t$. Note that $Q$ is a projection matrix, and since $\textrm{range}(W) \subset \textrm{range}(X)$, we have that $PQ = QP = P$ and $(I-P)(I-Q) = (I-Q)(I-P) = (I-Q)$.  In this notation, the fitted values are given by 
$$
Q y = \beta_{x|w} x + W \beta. 
$$
We now add and subtract terms as follows: 
$$
Q (y - P y + Py)  = \beta_{x|w} (x - Px + P x) + W \beta
$$
and rearrange to obtain: 
$$
Q (I-P) y  = \beta_{x|w} (I-P) x + P ( W \beta + \beta_{x|w} x - y). 
$$
We now apply $I-P$ to both sides, take the inner product with $x$ on both sides, use $Qx = x$, and rearrange to obtain 
\begin{subequations}
\label{first}
\begin{align}
\beta_{x|w} &= \frac{ \langle x,  (I-P) y\rangle } { \langle x, (I-P) x \rangle } \\
&=\frac{|y-\hat{y}|}{|x-\hat{x}|}\rho_{(x-\hat{x})(y-\hat{y})} \\
&=\frac{\sigma_y}{\sigma_x} \frac{\sqrt{1-R_{wy}^2}}{\sqrt{1-R_{wx}^2}}\rho_{(x-\hat{x})(y-\hat{y})}.
\end{align}
\end{subequations}
Combining \eqref{2b} and \eqref{first}, we obtain the desired result.
\hfill $\square$

\section{Proof of Proposition~\ref{prop2}} \label{ap:2}
In the setting of the proof of Proposition~\ref{formula}, consider \eqref{second}. 
Since 
$\rho_{(y-\hat{y})(x-\hat{x})} \in [-1,1]$, 
we obtain 
\begin{equation}\label{third}
\rho_{xy}\in [\xi_-, \xi_+], 
\qquad \textrm{where} \quad 
\xi_{\pm} = R_{wx}R_{wy}\rho_{\hat{x}\hat{y}}\pm \sqrt{1-R_{wx}^2}\sqrt{1-R_{wy}^2}.
\end{equation}  
Manipulating \eqref{third} to isolate $\rho_{\hat{x}\hat{y}}$ gives \eqref{nonline}. 

Conversely, if $n>p+2$,
there are sufficient degrees of freedom for $\rho_{(y-\hat{y})(x-\hat{x})}$ 
to take any desired value in $[-1,1]$.   
\hfill $\square$

\section{Proof of Proposition~\ref{prop3}} \label{ap:3}

We assume that the feasible set $\Omega$ is non-empty and is defined by the upper and lower inequality constraints in 
\eqref{e:intervalBoundsa}, 
\eqref{e:intervalBoundsb}, 
\eqref{e:intervalBoundsc}, and 
\eqref{nonline}, for eight inequality constraints in total.  
We do not allow $\alpha_-=\alpha_+$ in \eqref{nonline}, but we do allow $l_x=u_x$, $l_y=u_y$, and or $l_{\hat{x}\hat{y}}=u_{\hat{x}\hat{y}}$.  We say an inequality constraint is \emph{active} when it holds with equality.  
In what follows for $k=1,2,3$ we say 
\emph{exactly $k$ constraints are active} when exactly $k$ constraints are active and each active constraint is from a different line of 
\eqref{e:intervalBoundsa}, 
\eqref{e:intervalBoundsb}, 
\eqref{e:intervalBoundsc}, or 
\eqref{nonline}, {\textrm e.g.}, 
$l_x=R_{wx}^2=u_x$ counts as one constraint not two constraints.  
There is no ambiguity when zero constraints are active.  In three dimensions it is impossible or redundant to have four active constraints. 

Recalling our notation $l=\min_{\Omega}(\beta_{x|w})$ and $u=\max_{\Omega}(\beta_{x|w})$,  
we say a point $(R_{wx}^2,R_{wy}^2,\rho_{\hat{x}\hat{y}})$
is \emph{optimal} if it  satisfies  $\beta_{x|w}(R_{wx}^2,R_{wy}^2,\rho_{\hat{x}\hat{y}})=l$ 
or 
$\beta_{x|w}(R_{wx}^2,R_{wy}^2,\rho_{\hat{x}\hat{y}})=u$.  To prove Proposition \ref{prop3} we show that any optimal points in $\Omega$ must also be within the finite subset of points $S$.  $\Omega$ and $S$ depend on parameters $\{\rho_{xy},\sigma_y/\sigma_x;l_x,u_x,l_y,u_y,l_{\hat{x}\hat{y}},u_{\hat{x}\hat{y}}\}$ as described in Section \ref{meth}.  Recall, $\beta_{x|w}=\frac{\sigma_y}{\sigma_x}\frac{\rho_{xy}-R_{wx}R_{wy}\rho_{\hat{x}\hat{y}}}{1-R_{wx}^2}$.  
Note that
\[
\frac{\partial \beta_{x|w}}{\partial R_{wx}^2}=0\iff \frac{\partial \beta_{x|w}}{\partial R_{wx}}=0 \textrm{~and~} \frac{\partial \beta_{x|w}}{\partial R_{wy}^2}=0\iff \frac{\partial \beta_{x|w}}{\partial R_{wy}}=0\]
wherever $R_{wx}\neq 0$ and $R_{wy}\neq 0$.  To simplify algebra in what follows we compute $\frac{\partial \beta_{x|w}}{\partial R_{wx}}$ in place of $\frac{\partial \beta_{x|w}}{\partial R^2_{wx}}$ and $\frac{\partial \beta_{x|w}}{\partial R_{wy}}$ in place of $\frac{\partial \beta_{x|w}}{\partial R^2_{wy}}$.  Likewise, we infer the constancy of $\beta_{x|w}$ in $R_{wx}^2$ or $R_{wy}^2$ from its constancy in $R_{wx}$ or $R_{wy}$ respectfully.

Wherever no constraints are active $\frac{\partial \beta_{x|w}}{\partial \rho_{\hat{x}\hat{y}}}=\frac{\sigma_y}{\sigma_x}\frac{-R_{wx}R_{wy}}{1-R_{wx}^2}\neq 0$ and no optimal points exist.  If exactly one constraint is active, we have the following cases. 
If the constraint is from either \eqref{e:intervalBoundsa} or \eqref{e:intervalBoundsb}, 
then 
$\frac{\partial \beta_{x|w}}{\partial \rho_{\hat{x}\hat{y}}}=\frac{\sigma_y}{\sigma_x}\frac{-R_{wx}R_{wy}}{1-R_{wx}^2}\neq 0$. 
If that constraint is \eqref{e:intervalBoundsc}, then either 
$\frac{\partial \beta_{x|w}}{\partial R_{wy}}=\frac{\sigma_y}{\sigma_x}\frac{-R_{wx}\rho_{\hat{x}\hat{y}}}{1-R_{wx}^2}\neq 0$ 
or, if $\rho_{\hat{x}\hat{y}}=0$,  
$\beta_{x|w}$ is constant in $R_{wy}^2$. 
Finally, if that one constraint is \eqref{nonline}, then  
\begin{equation}
\label{firstb}
\beta_{x|w}=\pm \frac{\sigma_y}{\sigma_x}\sqrt{1-R_{wy}^2}/\sqrt{1-R_{wx}^2},
\end{equation} 
which is strictly monotonic in both $R_{wx}^2$ and $R_{wy}^2$ on both surfaces 
\begin{equation}
\label{surface}
\rho_{\hat{x}\hat{y}}=\frac{\rho_{xy}\pm\sqrt{1-R_{wx}^2}\sqrt{1-R_{wy}^2}}{R_{wx}R_{wy}}. 
\end{equation} 
We have thus far shown that two or more constraints must be active in order for a point to be optimal.

Suppose exactly two constraints are active.  
If those two constraints are \eqref{e:intervalBoundsa} and \eqref{e:intervalBoundsb} then $\beta_{x|w}=\frac{\sigma_y}{\sigma_x}\frac{\rho_{xy}-b_xb_y\rho_{\hat{x}\hat{y}}}{1-b_x^2}$ and either $\frac{\partial \beta_{x|w}}{\partial \rho_{\hat{x}\hat{y}}}\neq 0$ or $\beta_{x|w}$ is constant in $\rho_{\hat{x}\hat{y}}$.  
If those two constraints are \eqref{e:intervalBoundsa} and \eqref{e:intervalBoundsc} then $\beta_{x|w}=\frac{\sigma_y}{\sigma_x}\frac{\rho_{xy}-b_xR_{wy}b_{\hat{x}\hat{y}}}{1-b_x^2}$ and either $\frac{\partial \beta_{x|w}}{\partial R_{wy}}\neq 0$ or $\beta_{x|w}$ is constant in $R^2_{wy}$. 
If those two constraints are \eqref{e:intervalBoundsb} and \eqref{e:intervalBoundsc} then $\beta_{x|w}=\frac{\sigma_y}{\sigma_x}\frac{\rho_{xy}-R_{wx}b_yb_{\hat{x}\hat{y}}}{1-R_{wx}^2}$ 
and because 
$\frac{\partial \beta_{x|w}}{\partial R_{wx}}=\frac{\sigma_y}{\sigma_x}\frac{-b_yb_{\hat{x}\hat{y}}R_{wx}^2+2\rho_{xy}R_{wx}^2-b_yb_{\hat{x}\hat{y}}}{(1-R_{wx}^2)^2}$ optimal points require \[R_{wx}^2=q_\pm^2(-b_yb_{\hat{x}\hat{y}},2\rho_{xy},-b_yb_{\hat{x}\hat{y}})\] producing the options in line \eqref{firstset} of Proposition \ref{prop3}.  

We now consider exactly two active constraints and require one of them to be \eqref{nonline}.  
If those two constraints are \eqref{e:intervalBoundsa} and \eqref{nonline} then as in \eqref{firstb} we have $\beta_{x|w}$ strictly monotonic in $R_{wy}^2$.  
If those two constraints are \eqref{e:intervalBoundsb} and \eqref{nonline} then as in \eqref{firstb} we have $\beta_{x|w}$ strictly monotonic in $R_{wx}^2$.  
If those two constraints are \eqref{e:intervalBoundsc} and \eqref{nonline} then $b_{\hat{x}\hat{y}}=\frac{\rho_{xy}\pm\sqrt{1-R_{wx}^2}\sqrt{1-R_{wy}^2}}{R_{wx}R_{wy}}$, or written differently 
\begin{equation}
\label{secondb}
g(R_{wx},R_{wy}) := b_{\hat{x}\hat{y}}R_{wx}R_{wy}\pm\sqrt{1-R_{wx}^2}\sqrt{1-R_{wy}^2} 
= \rho_{xy}.
\end{equation}  
Along a curve $g(R_{wx},R_{wy})=\rho_{xy}$, via \eqref{firstb}, 
\[
h(R_{wx},R_{wy}):=\beta_{x|w}=\pm \frac{\sigma_y}{\sigma_x}\sqrt{1-R_{wy}^2}/\sqrt{1-R_{wx}^2}.
\]  
At an optimal point, for some real $\lambda$, we must have $\nabla h = \lambda \nabla g$ \citep[p. 31]{Noc}, implying equality between \[\frac{\partial g}{\partial R_{wx}}/\frac{\partial g}{\partial R_{wy}}=\frac{b_{\hat{x}\hat{y}}R_{wy}\pm 2R_{wx}(1+R_{wy}^2)^{1/2}(1-R_{wx}^2)^{-1/2}}{b_{\hat{x}\hat{y}}R_{wx}\pm 2R_{wy}(1+R_{wx}^2)^{1/2}(1-R_{wy}^2)^{-1/2}}\] and
\begin{align*}\frac{\partial h}{\partial R_{wx}}/\frac{\partial h}{\partial R_{wy}}&=\frac{R_{wx}(1-R_{wy}^2)^{1/2}(1-R_{wx}^2)^{-3/2}}{-R_{wy}(1-R_{wx}^2)^{-1/2}(1-R_{wy})^{-1/2}}\\&=\frac{R_{wx}(1-R_{wy}^2)}{-R_{wy}(1-R_{wx}^2)},\end{align*}
which in turn implies $R_{wx}=R_{wy}$.  Solving for $R^2=R_{wx}^2=R_{wy}^2$ within (\ref{secondb}) produces $R^2=\frac{\rho_{xy}+ 1}{b_{\hat{x}\hat{y}}+ 1}$ or  $R^2=\frac{\rho_{xy}- 1}{b_{\hat{x}\hat{y}}- 1}$ 
resulting in lines \eqref{secondset} and \eqref{secondsetb} of Proposition \ref{prop3}.  

Suppose exactly three constraints are active.  If those three constraints are \eqref{e:intervalBoundsa}, \eqref{e:intervalBoundsb}, and \eqref{e:intervalBoundsc} then any candidate point is of the form $(b_x^2,b_y^2,b_{\hat{x}\hat{y}})$ as in line \eqref{thirdset} of Proposition \ref{prop3}.  
If those three constraints are \eqref{e:intervalBoundsa}, \eqref{e:intervalBoundsb}, and \eqref{nonline} then via \eqref{surface} we have $\rho_{\hat{x}\hat{y}}=\frac{\rho_{xy}\pm\sqrt{1-b_x^2}\sqrt{1-b_y^2}}{b_xb_y}$ resulting in line \eqref{fourthset} of Proposition \ref{prop3}.  
If those three constraints are \eqref{e:intervalBoundsa}, \eqref{e:intervalBoundsc}, and \eqref{nonline} then via \eqref{surface} we have $b_{\hat{x}\hat{y}}=\frac{\rho_{xy}\pm\sqrt{1-b_x^2}\sqrt{1-R_{wy}^2}}{b_xR_{wy}}$, 
which after rearrangement and squaring gives \begin{equation}
\label{thirdb}
(b_x^2b^2_{\hat{x}\hat{y}}+1-b_x^2)R_{wy}^2-2b_xb_{\hat{x}\hat{y}}\rho_{xy}R_{wy}+(b_x^2-1+\rho_{xy}^2)=0.
\end{equation}  
Using the quadratic formula on \eqref{thirdb} results in line \eqref{fifthset} of Propositon \ref{prop3}.  If those three constraints are \eqref{e:intervalBoundsb}, \eqref{e:intervalBoundsc}, and \eqref{nonline} then likewise via \eqref{surface} we have 
$b_{\hat{x}\hat{y}}=\frac{\rho_{xy}\pm\sqrt{1-R_{wx}^2}\sqrt{1-b_y^2}}{R_{wx}b_y}$ 
and with analogous rearrangement, squaring, and use of the quadratic formula we solve for $R_{wx}$ and derive line \eqref{sixthset} of Proposition \ref{prop3}.

\end{document}